\documentclass[journal]{IEEEtran}
\usepackage{cite}
\ifCLASSINFOpdf
\else
\fi
\usepackage{graphicx}
\usepackage{amsmath}
\usepackage{subfig}
\usepackage{float}
\usepackage{multirow}
\usepackage{textcomp}
\usepackage{xcolor}
\usepackage{textcomp}
\usepackage{siunitx}

\begin{document}
%
\title{Heart Sound Segmentation using Bidirectional LSTMs with Attention}
%
%
%

\author{Tharindu~Fernando,~\IEEEmembership{Student Member,~IEEE,}
        Houman~Ghaemmaghami,
         Simon~Denman,~\IEEEmembership{Member,~IEEE,}
        Sridha~Sridharan,~\IEEEmembership{Life Senior Member,~IEEE,}
        Nayyar~Hussain,
        ~and~ Clinton~Fookes,~\IEEEmembership{Senior Member,~IEEE.}
     
\IEEEcompsocitemizethanks{\IEEEcompsocthanksitem T. Fernando, S.Denman, S. Sridharan and C. Fookes are with Speech and Audio Research Lab, SAIVT, Queensland University of Technology, Australia.\protect\\ H. Ghaemmaghami and N. Hussain are with M3DICINE Pty Ltd, Brisbane, Australia. }}

%
%

\markboth{Heart Sound Segmentation using Bidirectional LSTMs with Attention}%
{Fernando \MakeLowercase{\textit{et al.}}: journal of biomedical and health informatics}
%



\maketitle

\begin{abstract}
\textit{Objective: }This paper proposes a novel framework for the segmentation of phonocardiogram (PCG) signals into heart states, exploiting the temporal evolution of the PCG as well as considering the salient information that it provides for the detection of the heart state. \textit{Methods: }We propose the use of recurrent neural networks and exploit recent advancements in attention based learning to segment the PCG signal. This allows the network to identify the most salient aspects of the signal and disregard uninformative information. \textit{Results: }The proposed method attains state-of-the-art performance on multiple benchmarks including both human and animal heart recordings. Furthermore, we empirically analyse different feature combinations including envelop features, wavelet and Mel Frequency Cepstral Coefficients (MFCC), and provide quantitative measurements that explore the importance of different features in the proposed approach. \textit{Conclusion: }We demonstrate that a recurrent neural network coupled with attention mechanisms can effectively learn from irregular and noisy PCG recordings. Our analysis of different feature combinations shows that MFCC features and their derivatives offer the best performance compared to classical wavelet and envelop features. \textit{Significance: }Heart sound segmentation is a crucial pre-processing step for many diagnostic applications. The proposed method provides a cost effective alternative to labour extensive manual segmentation, and provides a more accurate segmentation than existing methods. As such, it can improve the performance of further analysis including the detection of murmurs and ejection clicks. The proposed method is also applicable for detection and segmentation of other one dimensional biomedical signals.

\end{abstract}


%
\IEEEpeerreviewmaketitle

\section{Introduction}

\IEEEPARstart{A}{mong} different diagnostic techniques, cardiac auscultation is one of the simplest and most cost effective methods for screening numerous heart conditions, including arrhythmia, valve disease and heart failure. However, as described in \cite{pease2001if,renna2019deep} heart sounds are difficult for human listeners to analyse due to faint sounds and significant events being closely spaced in time. Hence, computer aided heart sound analysis is becoming increasingly popular in healthcare as a cost effective alternative to manual monitoring which requires the extensive training of the human practitioners to master cardiac auscultation \cite{renna2019deep,ghaemmaghami2017automatic}.

A key component in computer aided heart sound analysis is the segmentation of the phonocardiogram (PCG) signal, as it allows the detection of the presence of extra sound components such as murmurs and ejection clicks in the PCG and enables further processing and analysis of individual components. 

\begin{figure}[htbp]
\centering
 {\includegraphics[width=\linewidth]{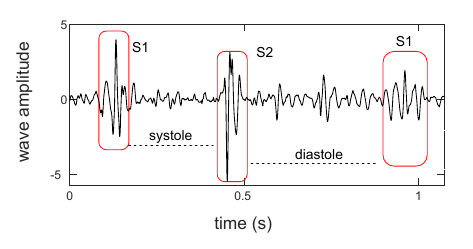}}
 \caption{Visualisation of heart states and systole / diastole intervals.}
\label{fig:states}
\end{figure}

In each heart cycle we observe the first heart sound (S1), caused due to the variation in blood pressure produced by the closure of the mitral and tricuspid valves and their vibrations; and the second heart sound (S2), generated by the closure of the aortic and pulmonary valves and their vibrations. The systole interval is the window between S1 and S2, and the diastole interval is from S2 to the beginning of S1 in the next heart cycle. Fig. \ref{fig:states} visually illustrates these states and intervals. Segmenting these S1 and S2 sounds in PCGs are challenging due to the accumulation of noise from the environment and the irregular sinus rhythm observed in the recordings \cite{messner2018heart, renna2019deep}.

The importance of accurate segmentation of heart sounds is revealed through the results of the 2016 PhysioNet Challenge on classification of abnormal heart sound recordings, where most of the proposed algorithms have considered a two step approach, first segmenting the heart sound recordings to states and then analysing the segmented signals for abnormal sounds. In those systems a significant performance boost is observed by augmenting the heart sound segmentation algorithm using advanced features and modelling techniques, but not by using sophisticated abnormal event classifiers \cite{renna2019deep}, which clearly exhibits the necessity of proper modelling within the segmentation algorithm. 

Motivated by the recent success of Recurrent Neural Networks (RNNs) when learning under noisy conditions \cite{dabre2018recurrent,fernando2018soft+, messner2018heart}, we exploit RNNs for the heart state segmentation task from a PCG. In particular, we employ an attention based learning framework to automatically understand the salient features in the current context, removing noisy and uninformative features from the observed PCG signal. Furthermore, we evaluate different feature combinations, ranging from classical wavelet \cite{springer2016logistic}, and envelop \cite{messner2018heart} features to MFCCs, deltas and delta-deltas. 

The main contributions of this work can be summarised as follows:

\begin{itemize}
\item We propose a novel a learning based approach for the segmentation of heart sounds which identifies the most salient features in noisy and irregular PCG signals.
\item We empirically evaluate different feature combinations, including MFCCs, wavelet entropy, power spectral density, Hilbert Envelop and Homomorphic Envelop features for the segmentation task.
\item We attain state-of-the-art results for multiple benchmarks including both human and animal heart sound databases. 
\item We interpret the model decisions using SHapley Additive exPlanation (SHAP) \cite{lundberg2017unified, lundberg2018explainable} feature importance distributions and demonstrate that $\Delta$ and $\Delta^2$ features are the most influential feature cues in the proposed method.   
\item We evaluate different architectural variants of the proposed framework, identifying crucial components of the architecture. 
\end{itemize}

Though deep learning models such as CNNs are extensively utilised in heart sound segmentation literature, to the best of our knowledge the only work to employ RNNs for the segmentation task is \cite{messner2018heart}. Hence the proposed work not only achieves state-of-the-art results on multiple benchmarks, but also significantly contributes to the biomedical engineering knowledge base in the area of heart sound segmentation.

\section{Related Work}

There have been numerous attempts to perform PCG segmentation. In several earlier works \cite{liang1997heart,ari2008robust,moukadem2013robust,sun2014automatic,huiying1997heart,castro2013heart} envelograms are extracted from the PCG signals and then a peak-picking \cite{andreev2003universal} algorithm is utilised to detect the principle heart sounds, S1 and S2. In \cite{liang1997heart,ari2008robust} the authors utilise energy based functions to extract envelograms while in \cite{sun2014automatic} the Hilbert transform is used. In \cite{huiying1997heart,castro2013heart} the authors employ a wavelet transform for the task. However, as pointed out in \cite{renna2019deep}, these methods lack robustness across different acquisition devices and show minimal resilience in the presence of noise. 

In a different line of work, methods such as \cite{naseri2013detection,kumar2006detection,vepa2009classification,gupta2007neural,chen2010intelligent,stasis2003using,sepehri2010novel} propose two stage approaches where a heart sound activity detection algorithm is first applied to identify possible segments for labelling. Then different features, including frequency domain features \cite{kumar2006detection}, and wavelets \cite{vepa2009classification} are extracted and classified into their respective states (i.e. S1, S2 and None). It is apparent that these methods are heavily reliant on the performance of the heart sound detection algorithm.  

Researches have also investigated the effect of temporal modelling for PCG segmentation. In \cite{springer2016logistic} the authors utilise a logistic regression Hidden Semi-Markov Model (HSMM) to predict the sequence of heart states. In \cite{oliveira2018adaptive} the authors propose an algorithm that evaluates sojourn time distribution parameters of the HSMM for each individual subject. Renna et. al \cite{renna2019deep} first encodes the features from the PCG signal through a CNN and then maps them sequentially through a HSMM. Most recently, in \cite{messner2018heart} the authors propose an event detection based approach where they utilise RNNs to classify the PCG signal, generating a prediction for each frame in the PCG. Their experimental evaluation suggests that RNN based temporal modelling is more robust when encountering noisy observations. 

In our proposed approach we leverage the merits of RNNs and attention mechanisms to filter out the salient features in the presence of noise and irregular heart signals, which allows us to obtain state-of-the-art results using the same framework for segmenting PCGs from both humans and animals. It should be noted that in contrast to the method of \cite{messner2018heart} which generates predictions for each frame in the given window, we provide a single classification to the entire window. We consider a frame shift of between 10ms to 20ms and temporally analyse how the PCG varies within that time window and provide a classification to the heart state. We utilise windows of small sizes to ensure that we do not observe multiple heart states within a single window.  

\section{Attention Based Learning Framework}

Motivated by the recent success of attention based RNNs \cite{dabre2018recurrent,fernando2018soft+, messner2018heart} for modelling noisy, irregular sequences, we employ these in our learning framework. 

Let $p$ denote a vector sampled from the PCG signal with a window from time instant $0$ to $T_{obs}$,
\begin{equation}
p=[X_0, \ldots, X_{T_{obs}}].
\end{equation}

Then we define a feature extraction function, $f$, which extracts features from observation $p$,
\begin{equation}
s=f(p).
\end{equation}
We employ bi-directional LSTMs to encode these feature vectors into a hidden representation,
 \begin{equation}
\begin{split}
\overrightarrow{h_{t}}= LSTM_{fwd}(s_t, \overrightarrow{h_{t-1}}), \\
\overleftarrow{h_{t}}= LSTM_{bwd}(s_t, \overleftarrow{h_{t+1}}),
\end{split}
\end{equation}
where the forward pass, $\overrightarrow{h_{t}}$, of the LSTM reads from feature elements $0$ to $T_{obs}$ while in the backward pass, $\overleftarrow{h_{t}}$, reads from $T_{obs}$ to $0$ and concatenates the forward and backward vectors to generate a single vector, 
\begin{equation}
\overleftrightarrow{h_t}= [\overrightarrow{h_t} ; \overleftarrow{h_{t}}].
\end{equation}
While utilising all hidden states corresponding to frames in the PCG generally tends to fully capture the attributes within the given window, due to the presence of noise and irregularities in the signal, not all hidden states equally contribute to the final classification of the heart state. Hence we apply a soft attention mechanism that automatically learns to pay varying levels of attention to the elements in $\overleftrightarrow{h_t}$ when generating the final decision. Specifically, we pass the elements in $\overleftrightarrow{h_t}$ through a single layer MLP  \cite{rumelhart1985learning} and obtain a representation, $v_t$, using,
\begin{equation}
v_t=\mathrm{tanh}(W_h\overleftrightarrow{h_t} + b_h),
\end{equation} 
where the weights and bias of the MLP are denoted by $W_h$ and $b_h$, respectively. Then using a context vector, $\hat{v}$, we measure the importance of each element in $\overleftrightarrow{h_t}$ in terms of the similarity between $\overleftrightarrow{h_t}$ and $\hat{v}$. As in \cite{yang2016hierarchical} we randomly initialise $\hat{v}$ and jointly learn it in the training process. Using a softmax function we normalise the similarity score values, 

\begin{equation}
\beta_t = \dfrac{\mathrm{exp}([v_t]^\top \hat{v})}{\sum_{t}\mathrm{exp}([v_t]^\top \hat{v})}.
\label{eq:attention}
\end{equation}

We multiply each element in $\overleftrightarrow{h_t}$ by the respective score values, generating an augmented representation, 
\begin{equation}
q=\sum_{t}\beta_t \overleftrightarrow{h_t}.
\label{eq:att}
\end{equation}
As the final step, the respective heart state labels are regressed through an MLP such that, 
\begin{equation}
\eta = \mathrm{ReLu}(W_{\eta}q + b_{\eta}),
\end{equation}
where $W_{\eta}$ and $b_{\eta}$ are the weights and bias of the MLP. Note that instead of softmax classification we perform a regression task where in the ground truth we denote heart state S1 with $1$ and state S2 with $-1$. This generates an approximate sine wave which is more convenient for post-processing tasks using the model predictions, including threshold and signal clean up  operations which can consider the symmetry of the labels in contrast to using categorical labels. 

\section{Experiments}
\subsection{Datasets}
\subsubsection{PhysioNet/ CinC Challenge 2016 (PCC)}
As the first heart sound database we use the heart sound recordings from the 2016 PhysioNet/ CinC Challenge \cite{liu2016open}, which is composed of data collected by different research groups, obtained under clinical and non-clinical settings. Hence it contains numerous variations in recording hardware, data quality and patient type, generating a challenging setting for evaluation. The main objective of the challenge is to automatically detect abnormal heart sounds including murmurs and other pathological conditions. Apart from these annotations, the challenge dataset provides heart sound states (S1, systole, S2, diastole) which are generated with the LR-HSMM algorithm \cite{springer2016logistic} and manually corrected. 

Similar to \cite{messner2018heart}, due to the unavailability of ground truth for the test set of PCC, we randomly selected 764 recordings from the `training-a', `training-b' and `training-e' subsets as the test set. The rest of the dataset is divided into 210 recordings for validation and 1900 recordings for training. 

\subsubsection{M3dicine Human Heart Sound Database (M3-Hu)}
This dataset consists of 170 digital heart sound recordings collected from male and female subjects with healthy hearts using Stethee\textregistered{}\footnote{https://www.stethee.com/} (a wireless electronic stethoscope). Each recording has a standard length of 20 seconds as captured by Stethee\textregistered{}. The recordings have been manually annotated by trained physicians and two cardiologists.

\subsubsection{M3dicine Animal Heart Sound Database (M3-An)}
This dataset consists of 105 digital heart sound recordings collected from a wide range of animals using Stethee\textregistered{} and with the help of a vet. Each recording has a standard length of 20 seconds as captured by Stethee\textregistered. The recordings have been manually annotated by trained vets. The animals in the dataset include large dog breeds, small dogs, birds (cockatoo, amazonian parrot, chicken), domestic cats, guinnipigs, horses, cattle and swine. The variety in the heart muscle of these animals and their varying heart rates (minimum of 24 BPM for a horse and maximum of 315 BPM for a parrot) poses interesting challenges for the task of heart sound segmentation. In addition, the dataset reflects real-life noisy scenarios. This is because animals are often anxious when examined and thus their movement generates additional unwanted noise.

\subsection{Feature Extraction and Data Augmentation}

In the proposed work we utilise MFCCs \cite{imai1983cepstral}, Deltas of MFCCs ($\Delta$) and Delta-Deltas ($\Delta^2$) as the features of the input PCG signal. We resample the heart sound recordings to a sampling frequency of 1.6 kHz and similar to \cite{messner2018heart} we pre-process recordings using a Hamming window with a window size of 80 ms and a frame shift of 20 ms for humans and 10 ms for animals. 

As spectral features we use 6 static MFCC coefficients, 6 $\Delta$ coefficients and 6 $\Delta^2$ coefficients.  We use 6 Mel bands within the range 30-300 Hz. 

Additionally, in Sec. \ref{sec:classical_features} we present evaluations when the proposed model is trained using features including: Homomorphic Envelograms (HoE), Hilbert Envelograms (HiE), Wavelet Envelop (WE) and Power Spectral Density (PSD). We strictly adhere to the feature extraction setup in \cite{springer2016logistic} for the extraction of these features. 

Similar to \cite{messner2018heart} we perform data augmentation through injection of noise. We inject zero mean 15 dB Gaussian noise to the inputs. 

\subsection{Evaluation Metrics}

We evaluate the performance using the following five metrics: positive predictive value (PPV), Sensitivity (Se), Specificity (Spe), Accuracy (Acc) and F1. Let TP denote the heart states that are correctly classified by the proposed system. FP denotes the states that are indicated by the system that are not present in the ground truth, FN denotes states that are present in the ground truth but not detected by the system and TN are the actual none-states correctly classified by the system. Then,
\begin{equation}
PPv = \frac{TP}{TP + FP},
\end{equation}
\begin{equation}
Se = \frac{TP}{TP + FN},
\end{equation}
\begin{equation}
Spe = \frac{TN}{TN + FP},
\end{equation}
\begin{equation}
Acc = \frac{TP + TN}{TP + FP + TN + FN},
\end{equation}
\begin{equation}
F_1 = 2 \frac{PPv . Se}{ PPv + Se}.
\end{equation}

\subsection{Implementation Details}
\begin{figure}[htbp]
\includegraphics[width=\linewidth]{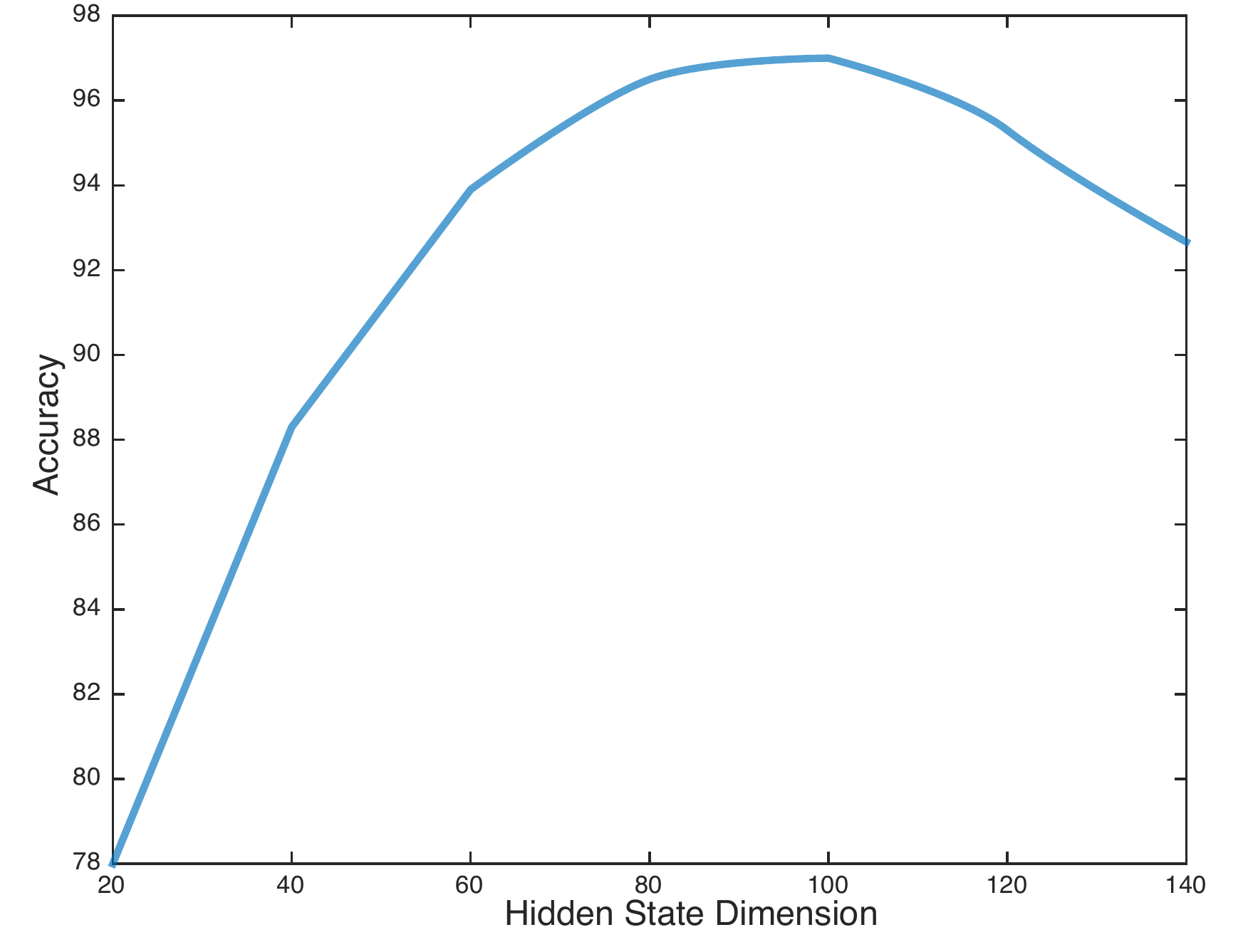}
\caption{Hyper-parameter Evaluation: We evaluate the hidden state dimension of the encoder LSTMs experimentally using the validation set of M3-Hu dataset, holding the rest of the parameters constant. As 80 hidden units provides best accuracy we set the hidden dimension as 80. }
\label{fig:hyp}
\end{figure}

The hidden state dimension of the encoder LSTMs is evaluated experimentally and is set to 80 units. Please refer to Fig. \ref{fig:hyp} for this evaluation. We use Keras \cite{chollet2015keras} with the Theano \cite{bergstra2010theano} backend for our implementations and trained the models using mean square error as our objective and the Adam \cite{kingma2014adam} optimiser, using a mini-batch size of 32 and a learning rate of 0.002 for 30 epochs and set the learning rate to be 0.0002 for another 70 epochs. 

\subsection{Results}

An evaluation of the proposed system together with the current state-of-the-art frameworks for the PCC \cite{liu2016open}, M3-Hu and M3-An datasets is presented in Tab. \ref{tab:all_datasets}. MFCC derived features (static, deltas, and delta-deltas) are used in this evaluation.In order to illustrate the robustness of the proposed system we randomly initialise the network parameters and ran each experiment 10 times, and report the average metric values along with the standard deviation.

\begin{figure*}[htbp]
\centering
 \subfloat[][Proposed]{\includegraphics[width=.16\textwidth]{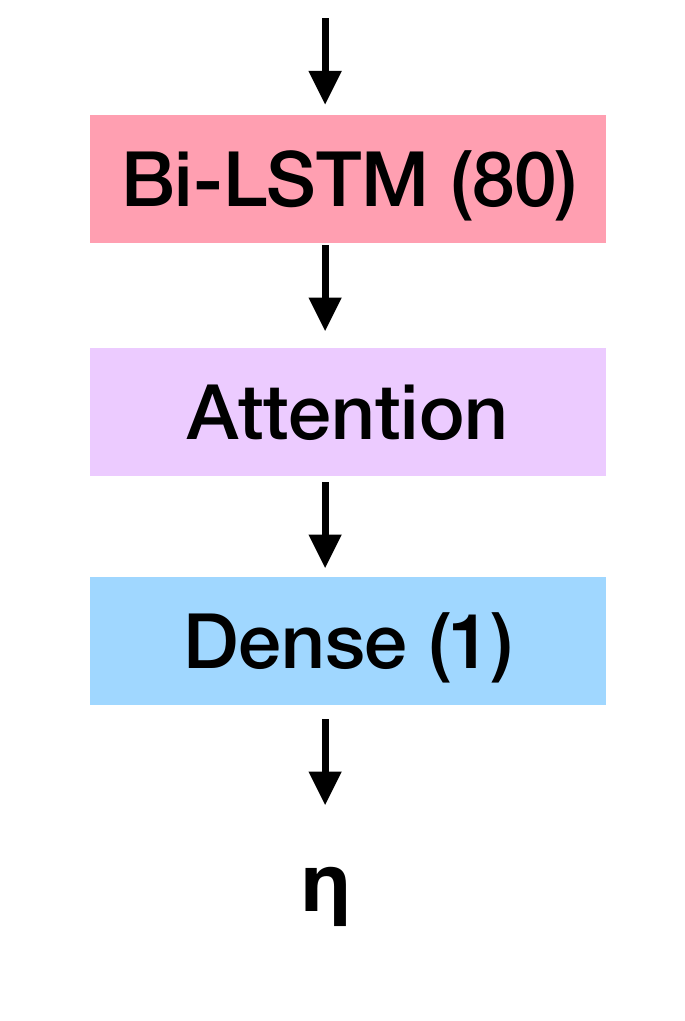}}
 \subfloat[][LSTM]{\includegraphics[width=.17\textwidth]{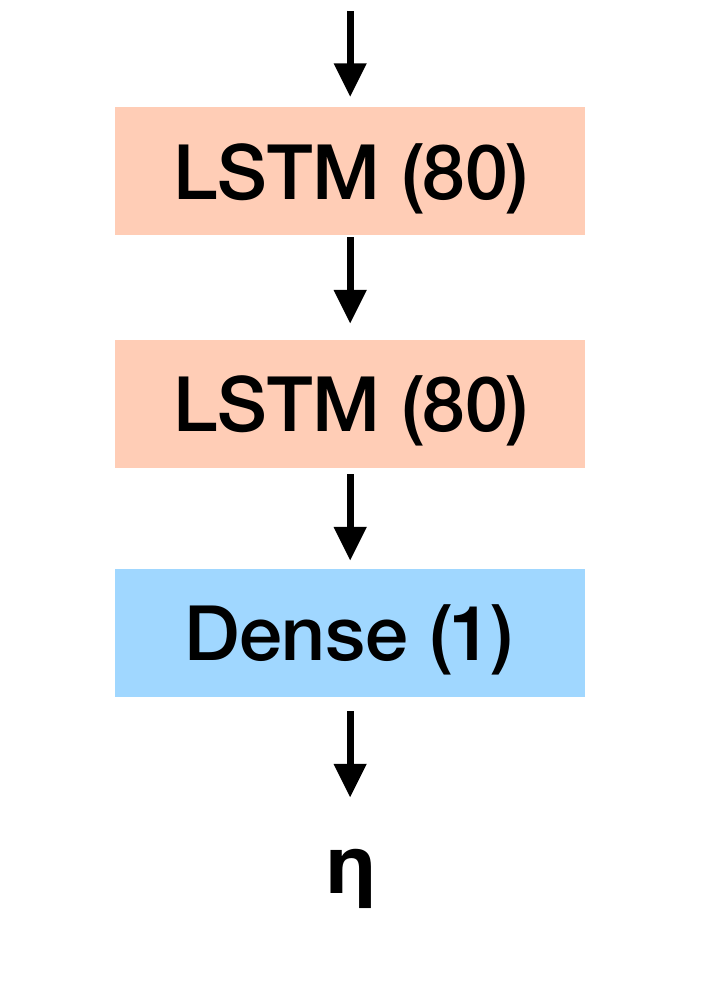}}
  \subfloat[][Bi-LSTM]{\includegraphics[width=.15\textwidth]{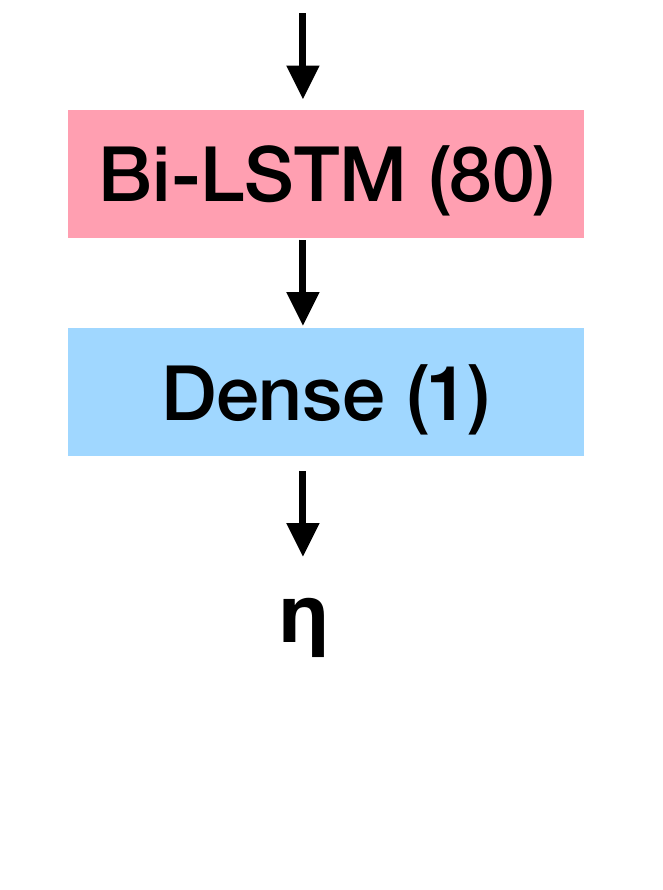}}
 \subfloat[][CNN]{\includegraphics[width=.175\textwidth]{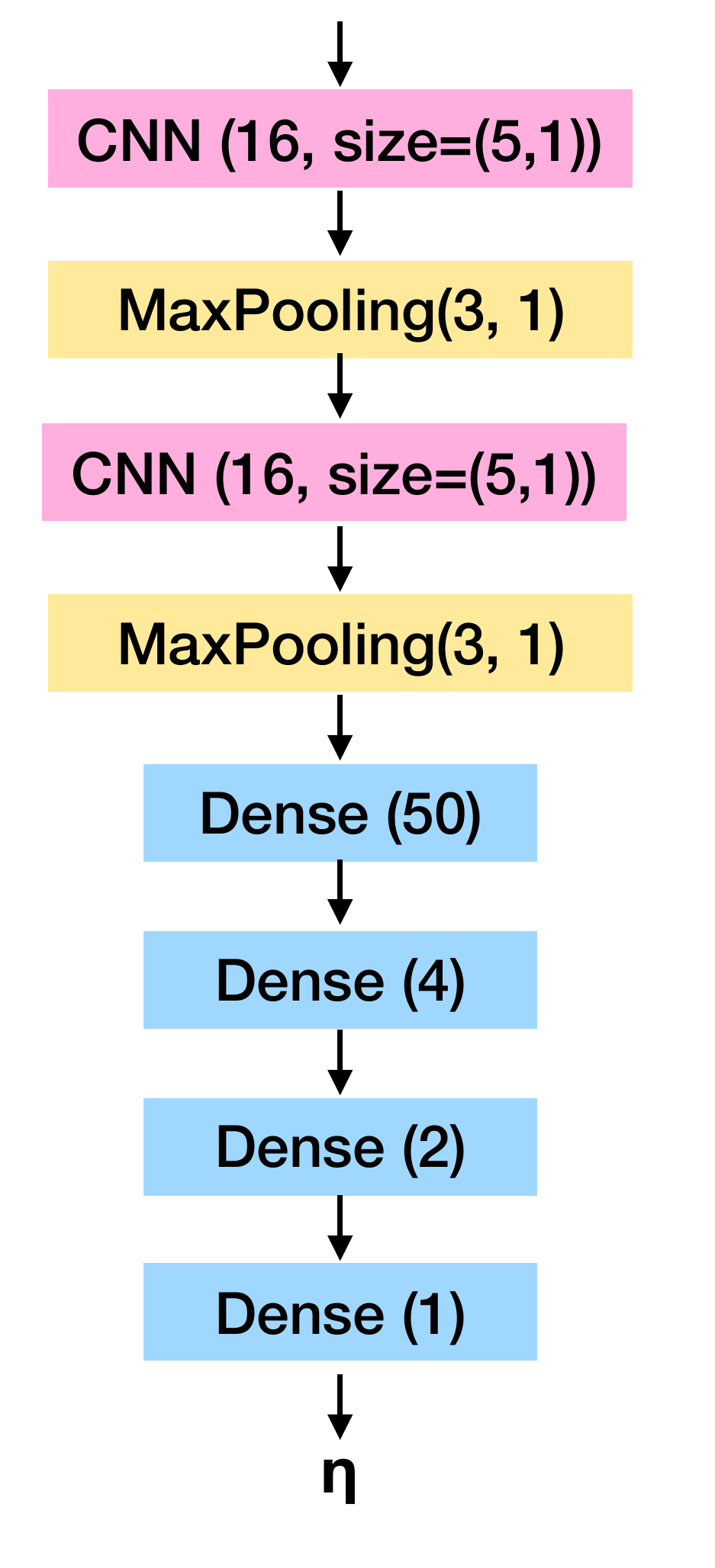}}
\caption{Model Architectures of the (a) proposed model along with (b) 2-layer LSTM model, (c) bi-directional LSTM (Bi-LSTM) model and (d) vanilla CNN model.}
\label{fig:models}
\end{figure*}

As baseline models we utilise the GRNN method of \cite{messner2018heart}, the DNN based method of \cite{chen2017s1}, a 2-layer LSTM model, a bi-directional LSTM (bi-LSTM) model and a vanilla CNN model. For the LSTM model we utilise a 2 layer LSTM with 80 hidden units, and for the bi-LSTM we use 80 hidden units for forward and backward layers. For the CNN model we use 2 blocks that each contain a convolutional layer with 16 kernels and max-pooling, followed by 4 dense layers. The architectures of proposed model and LSTM, bi-LSTM and CNN baselines are visually illustrated in Fig. \ref{fig:models}. 

Due to the unavailability of public implementations of GRNN  \cite{messner2018heart} and DNN \cite{chen2017s1} methods , we report the results of our own implementation of those algorithms. 

\begin{table*}[htbp]
\centering
\resizebox{0.75\linewidth}{!} {
\begin{tabular}{|c|c|c|c|c|c|c|}
\hline
Dataset                & Method   & PPv & Se & Spe & Acc & F1 \\ \hline
\multirow{6}{*}{PCC}   
		
		      & DNN \cite{chen2017s1}     & \SI{88.4}{} \textpm\ \SI{0.42}{}    &  \SI{89.8}{} \textpm\ \SI{0.23}{}  & \SI{89.5}{} \textpm\ \SI{0.21}{}     & \SI{89.2}{} \textpm\ \SI{0.81}{}    &  \SI{89.29}{} \textpm\ \SI{0.81}{}   \\ \cline{2-7} 
                       & CNN      &  \SI{94.2}{} \textpm\ \SI{0.35}{}   & \SI{95.0}{} \textpm\ \SI{0.12}{}   &  \SI{93.4}{} \textpm\ \SI{0.81}{}   & \SI{93.1}{} \textpm\ \SI{0.39}{}     & \SI{94.70}{} \textpm\ \SI{0.69}{}   \\ \cline{2-7} 
                       & GRNN \cite{messner2018heart}    & \SI{94.2}{} \textpm\ \SI{0.31}{}    & \SI{95.1}{} \textpm\ \SI{0.25}{}    & \SI{95.0}{} \textpm\ \SI{0.46}{}     &  \SI{93.3}{} \textpm\ \SI{0.81}{}   & \SI{94.75}{} \textpm\ \SI{0.25}{}   \\ \cline{2-7}
                       & LSTM     &   \SI{94.5}{} \textpm\ \SI{0.35}{}  & \SI{95.0}{} \textpm\ \SI{0.39}{}   &   \SI{95.1}{} \textpm\ \SI{0.21}{}  &   \SI{93.5}{} \textpm\ \SI{0.52}{}  & \SI{94.75}{} \textpm\ \SI{0.39}{}   \\ \cline{2-7} 
                       & Bi-LSTM  &   \SI{95.0}{} \textpm\ \SI{0.56}{}  &  \SI{96.4}{} \textpm\ \SI{0.15}{}  & \SI{96.8}{} \textpm\ \SI{0.32}{}    &  \SI{95.1}{} \textpm\ \SI{0.42}{}   & \SI{95.75}{} \textpm\ \SI{0.22}{}   \\ \cline{2-7} 
                       & Proposed & \textbf{\SI{96.3}{} \textpm\ \SI{0.42}{}}    & \textbf{\SI{97.2}{} \textpm\ \SI{0.19}{}}   & \textbf{\SI{97.5}{} \textpm\ \SI{0.20}{}}    & \textbf{\SI{96.9}{} \textpm\ \SI{0.13}{}}    & \textbf{\SI{96.70}{} \textpm\ \SI{0.17}{}}   \\ \hline \hline
                       
\multirow{6}{*}{M3-Hu} 
		      
                       & DNN \cite{chen2017s1}     & \SI{75.3}{} \textpm\ \SI{0.81}{}      & \SI{88.4}{} \textpm\ \SI{0.31}{}      & \SI{93.0}{} \textpm\ \SI{0.15}{}      & \SI{90.1}{} \textpm\ \SI{0.33}{}     & \SI{81.40}{} \textpm\ \SI{0.15}{}    \\ \cline{2-7}
                       & CNN      & \SI{87.2}{} \textpm\ \SI{0.61}{}     & \SI{93.2}{} \textpm\ \SI{0.33}{}     & \SI{95.9}{} \textpm\ \SI{0.44}{}     &  \SI{94.8}{} \textpm\ \SI{0.19}{}    & \SI{90.40}{} \textpm\ \SI{0.18}{}    \\ \cline{2-7}  
                       & GRNN \cite{messner2018heart}    & \SI{87.8}{} \textpm\ \SI{0.75}{}     & \SI{92.8}{} \textpm\ \SI{0.67}{}    & \SI{96.4}{} \textpm\ \SI{0.49}{}      & \SI{95.0}{} \textpm\ \SI{0.34}{}     & \SI{90.47}{} \textpm\ \SI{0.27}{}    \\ \cline{2-7} 
                       & LSTM     & \SI{88.2}{} \textpm\ \SI{0.59}{}     & \SI{93.6}{} \textpm\ \SI{0.56}{}    &   \SI{96.7}{} \textpm\ \SI{0.38}{}   &  \SI{95.1}{} \textpm\ \SI{0.48}{}    & \SI{90.82}{} \textpm\ \SI{0.22}{}    \\ \cline{2-7} 
                       & Bi-LSTM  & \SI{90.2}{} \textpm\ \SI{0.43}{}      & \SI{93.8}{} \textpm\ \SI{0.69}{}    & \SI{95.8}{} \textpm\ \SI{0.22}{}      & \SI{95.4}{} \textpm\ \SI{0.41}{}     &   \SI{92.32}{} \textpm\ \SI{0.49}{}   \\ \cline{2-7} 
                       & Proposed &  \textbf{\SI{93.1}{} \textpm\ \SI{0.22}{}}   & \textbf{\SI{96.7}{} \textpm\ \SI{0.24}{}}   &  \textbf{\SI{96.7}{} \textpm\ \SI{0.12}{}}    & \textbf{\SI{97.1}{} \textpm\ \SI{0.32}{}}    &\textbf{ \SI{94.70}{} \textpm\ \SI{0.15}{}}    \\ \hline  \hline
                       
\multirow{6}{*}{M3-An} 
		       & DNN  \cite{chen2017s1}    &  \SI{72.5}{} \textpm\ \SI{0.23}{}    & \SI{72.8}{} \textpm\ \SI{0.34}{}   &  \SI{95.3}{} \textpm\ \SI{0.54}{}   & \SI{91.8}{} \textpm\ \SI{0.12}{}    & \SI{72.81}{} \textpm\ \SI{0.45}{}    \\ \cline{2-7} 
		       & CNN      & \SI{85.6}{} \textpm\ \SI{0.20}{}    & \SI{92.0}{} \textpm\ \SI{0.12}{}   & \SI{94.8}{} \textpm\ \SI{0.56}{}    & \SI{94.1}{} \textpm\ \SI{0.27}{}    & \SI{88.74}{} \textpm\ \SI{0.43}{}   \\ \cline{2-7}
                       & GRNN \cite{messner2018heart}    & \SI{86.3}{} \textpm\ \SI{0.37}{}    &  \SI{92.2}{} \textpm\ \SI{0.42}{}  & \SI{94.6}{} \textpm\ \SI{0.22}{}     & \SI{94.5}{} \textpm\ \SI{0.35}{}    &  \SI{89.23}{} \textpm\ \SI{0.52}{}   \\ \cline{2-7} 
                       & LSTM     &  \SI{86.6}{} \textpm\ \SI{0.24}{}    & \SI{92.2}{} \textpm\ \SI{0.39}{}   &  \SI{95.1}{} \textpm\ \SI{0.19}{}   & \SI{94.6}{} \textpm\ \SI{0.15}{}     &  \SI{89.37}{} \textpm\ \SI{0.52}{}    \\ \cline{2-7} 
                       & Bi-LSTM  & \SI{87.2}{} \textpm\ \SI{0.14}{}    &  \SI{94.0}{} \textpm\ \SI{0.26}{}  &  \SI{95.8}{} \textpm\ \SI{0.75}{}   & \SI{95.3}{} \textpm\ \SI{0.45}{}     &  \SI{90.55}{} \textpm\ \SI{0.49}{}  \\ \cline{2-7}
                       & Proposed &  \textbf{\SI{91.1}{} \textpm\ \SI{0.17}{}}   & \textbf{\SI{95.4}{} \textpm\ \SI{0.29}{}}   &  \textbf{\SI{96.2}{} \textpm\ \SI{0.28}{}}   & \textbf{\SI{96.0}{} \textpm\ \SI{0.17}{}}    &\textbf{\SI{93.25}{} \textpm\ \SI{0.28}{}}    \\ \hline
\end{tabular}}
\caption{Evaluation Results on PCC \cite{liu2016open} M3-Hu and M3-An datasets.}
\label{tab:all_datasets}
\end{table*}

When analysing the results it is clear that RNN based systems (i.e LSTM, bi-LSTM, GRNN and Proposed) outperform the CNN and DNN based counterparts, denoting the importance of temporal modelling. Furthermore, when comparing the LSTM model with the bi-LSTM model we observe the bi-directional propagation in the bi-LSTM model has contributed to a slight performance boost. Due to the random initialisation of the network weights at each of the 10 trials we observe a slight fluctuation in performance for all models. Despite of these minor fluctuations we observe that the proposed method has always been able to converge to a state which produces higher accuracy than the baseline models. We believe that this significant increase in performance results from the utilisation of the proposed attention mechanism, denoting the importance of selecting the most salient features in the presence of noise. Furthermore, we would like to point out the fact that the GRNN method utilises spectrogram and envelope \cite{springer2016logistic} features in addition to the MFCC, $\Delta$ and $\Delta^2$ features that we utilise, which incurs additional computation cost but results in lesser performance compared to the proposed method. 

\subsection{Comparison with Classical Features}
\label{sec:classical_features}
In Tab. \ref{tab:classical_feat} we present an evaluation of the proposed model when trained with classical HoE, HiE, WE, PSD features and MFCC, $\Delta$ and $\Delta^2$ features and combinations of these. In this experiment we use the M3-Hu dataset. 

\begin{table*}[htbp]
\centering
\resizebox{.85\textwidth}{!}{
\begin{tabular}{|c|c|c|c|c|c|}
\hline
Method                                    & PPv & Se & Spe & Acc & F1 \\ \hline
HoE                                       &   \SI{87.4}{} \textpm\ \SI{0.72}{} &  \SI{81.0}{} \textpm\ \SI{0.98}{} & \SI{70.2} {} \textpm\ \SI{0.73}{}& \SI{77.1} {} \textpm\ \SI{0.13}{} & \SI{80.85} {} \textpm\ \SI{0.14}{}  \\ \hline
HiE                                       &     \SI{81.2}{} \textpm\ \SI{0.78}{} & \SI{83.1}{} \textpm\ \SI{0.92}{} & \SI{80.2}{} \textpm\ \SI{0.55}{} & \SI{79.2}{} \textpm\ \SI{0.23}{}    &  \SI{82.20}{} \textpm\ \SI{0.24}{}  \\ \hline
WE                                        &     \SI{75.8}{} \textpm\ \SI{0.75}{} & \SI{79.4}{} \textpm\ \SI{0.72}{} & \SI{67.8}{} \textpm\ \SI{0.29}{} & \SI{73.7}{} \textpm\ \SI{0.56}{}     &  \SI{78.55}{} \textpm\ \SI{0.22}{}  \\ \hline
PSD                                       &     \SI{73.7}{} \textpm\ \SI{0.61}{} & \SI{71.2}{} \textpm\ \SI{0.54}{} & \SI{67.8}{} \textpm\ \SI{0.91}{} & \SI{74.7}{} \textpm\ \SI{0.45}{}     & \SI{72.46}{} \textpm\ \SI{0.26}{}   \\ \hline
MFCC                                      &     \SI{86.1}{} \textpm\ \SI{0.55}{} & \SI{90.2}{} \textpm\ \SI{0.48}{} & \SI{81.5}{} \textpm\ \SI{0.36}{} & \SI{85.2 }{} \textpm\ \SI{0.62}{}    &    \SI{87.80}{} \textpm\ \SI{0.45}{} \\ \hline
$\Delta$                                     &     \SI{87.2}{} \textpm\ \SI{0.91}{} & \SI{81.2}{} \textpm\ \SI{0.67}{} & \SI{82.4}{} \textpm\ \SI{0.22}{} & \SI{87.8}{} \textpm\ \SI{0.31}{}     &   \SI{85.31}{} \textpm\ \SI{0.61}{} \\ \hline
$\Delta^2$                                   &     \SI{81.8}{} \textpm\ \SI{0.78}{} & \SI{82.0}{} \textpm\ \SI{0.42}{} & \SI{72.7}{} \textpm\ \SI{0.68}{} & \SI{78.1}{} \textpm\ \SI{0.25}{}     & \SI{82.44}{} \textpm\ \SI{0.53}{}   \\ \hline
HoE + WE                                  &     \SI{82.5}{} \textpm\ \SI{0.71}{} & \SI{82.3}{} \textpm\ \SI{0.67}{} & \SI{78.7}{} \textpm\ \SI{0.49}{} & \SI{79.2}{} \textpm\ \SI{0.34}{}     &  \SI{82.40}{} \textpm\ \SI{0.27}{}   \\ \hline
HiE+ WE                                   &     \SI{85.6}{} \textpm\ \SI{0.65}{} & \SI{86.2}{} \textpm\ \SI{0.52}{} & \SI{84.1}{} \textpm\ \SI{0.43}{} & \SI{81.2}{} \textpm\ \SI{0.51}{}     & \SI{85.79}{} \textpm\ \SI{0.53}{}    \\ \hline
HoE+ HiE+ WE+ PSD                         &    \SI{83.4}{} \textpm\ \SI{0.42}{} & \SI{78.2}{} \textpm\ \SI{0.43}{} & \SI{78.9}{} \textpm\ \SI{0.23}{} & \SI{80.5}{} \textpm\ \SI{0.41}{}      &  \SI{81.60}{} \textpm\ \SI{0.61}{}  \\ \hline
WE + PSD + MFCC                           &     \SI{84.3}{} \textpm\ \SI{0.65}{} & \SI{85.2}{} \textpm\ \SI{0.29}{} & \SI{85.1}{} \textpm\ \SI{0.36}{} & \SI{84.2}{} \textpm\ \SI{0.29}{}     & \SI{85.44}{} \textpm\ \SI{0.78}{}    \\ \hline
WE+ HiE + MFCC + $\Delta$                    &     \SI{90.1}{} \textpm\ \SI{0.82}{} & \SI{93.8}{} \textpm\ \SI{0.52}{} & \SI{92.9}{} \textpm\ \SI{0.66}{} & \SI{92.8}{} \textpm\ \SI{0.12}{}     &  \SI{91.53}{} \textpm\ \SI{0.52}{}  \\ \hline
WE + MFCC + $\Delta$ + PSD                   &     \SI{90.3}{} \textpm\ \SI{0.81}{} & \SI{92.6}{} \textpm\ \SI{0.28}{} & \SI{92.8}{} \textpm\ \SI{0.59}{} & \SI{93.1}{} \textpm\ \SI{0.29}{}     & \SI{91.50}{} \textpm\ \SI{0.32}{}   \\ \hline
WE+ HoE+ HiE + PSD+ MFCC + $\Delta$+ $\Delta^2$  &  \SI{91.1}{} \textpm\ \SI{0.76}{}   & \SI{94.9}{} \textpm\ \SI{0.23}{}   &   \SI{95.9}{} \textpm\ \SI{0.51}{}    &   \SI{96.3}{} \textpm\ \SI{0.32}{}   &    \SI{93.25}{} \textpm\ \SI{0.24}{}    \\ \hline
MFCC + $\Delta$+ $\Delta^2$ &  \textbf{\SI{93.1}{} \textpm\ \SI{0.22}{}}   & \textbf{\SI{96.7}{} \textpm\ \SI{0.24}{}}   &  \textbf{\SI{96.7}{} \textpm\ \SI{0.12}{}}    & \textbf{\SI{97.1}{} \textpm\ \SI{0.32}{}}    &\textbf{ \SI{94.70}{} \textpm\ \SI{0.15}{}} \\ \hline

\end{tabular}}
\caption{Evaluation Results on M3-Hu when proposed method is trained with classical HoE, HiE, WE, PSD features and MFCC, $\Delta$ and $\Delta^2$ features and their combinations }
\label{tab:classical_feat}
\end{table*}

We observe a strong performance solely with MFCC, $\Delta$ and $\Delta^2$ features and only a slight performance increase when adding envelope features. One interesting observation is that the $\Delta$ feature has attained good performance compared to using other features individually. We further explore this observation in Sec. \ref{sec:interpritation}.

\subsection{Model Interpretation}
\label{sec:interpritation}

We utilise the SHapley Additive exPlanation (SHAP) \cite{lundberg2017unified, lundberg2018explainable} framework for interpreting model outputs. The SHAP values denote the change in the expected model predictions when the model is conditioned on a particular feature. They explain what the model would have predicted if the model did not know any information regarding the other features \cite{lundberg2017unified}. Hence it provides a quantitative measure regarding the feature importance. We utilise two examples from the M3-Hu dataset in our analysis. 

\begin{figure*}[htbp]
\centering
 {\includegraphics[width=.8\textwidth]{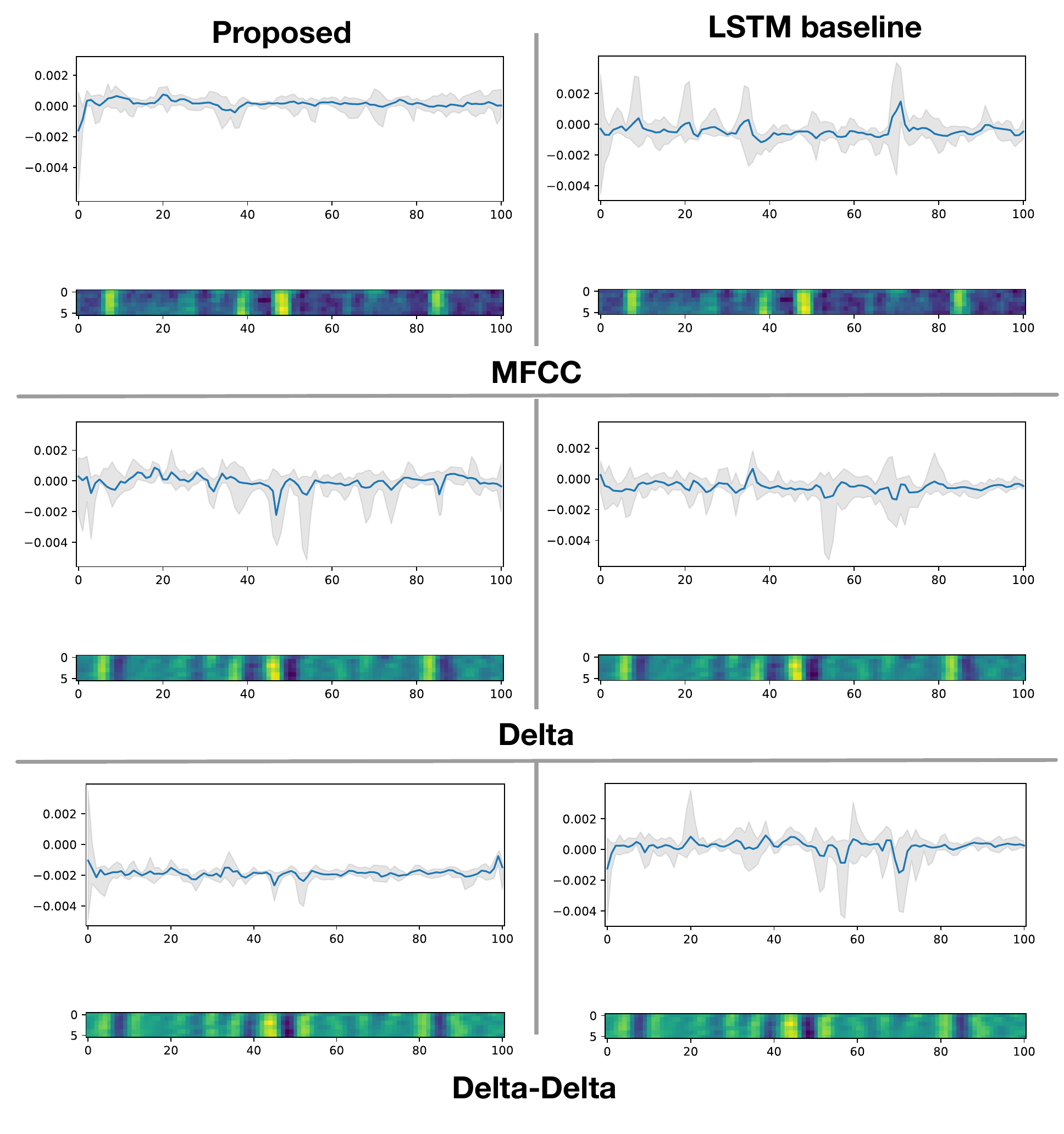}}
 \caption{SHAP \cite{lundberg2017unified, lundberg2018explainable} values denoting the change in the expected model predictions when the model is expected to generate a prediction of -1. In rows we indicate the MFCC, $\Delta$ and $\Delta^2$ features and in first and second columns we indicate the proposed method and LSTM baselines. As MFCC, $\Delta$ and $\Delta^2$ features contain 6 coefficients, we visualise the mean value in blue and the distribution in grey. We also illustrate the heat maps of the features. We observe that the $\Delta$ feature has impacted more towards generating accurate predictions from the proposed model compared to MFCC and $\Delta^2$ features while fluctuations in model SHAP values are observed in baseline LSTM model denoting that model is unsure about the output that it produces.}
\label{fig:shap_1}
\end{figure*}

\begin{figure*}[htpb]
\centering
 {\includegraphics[width=.8\textwidth]{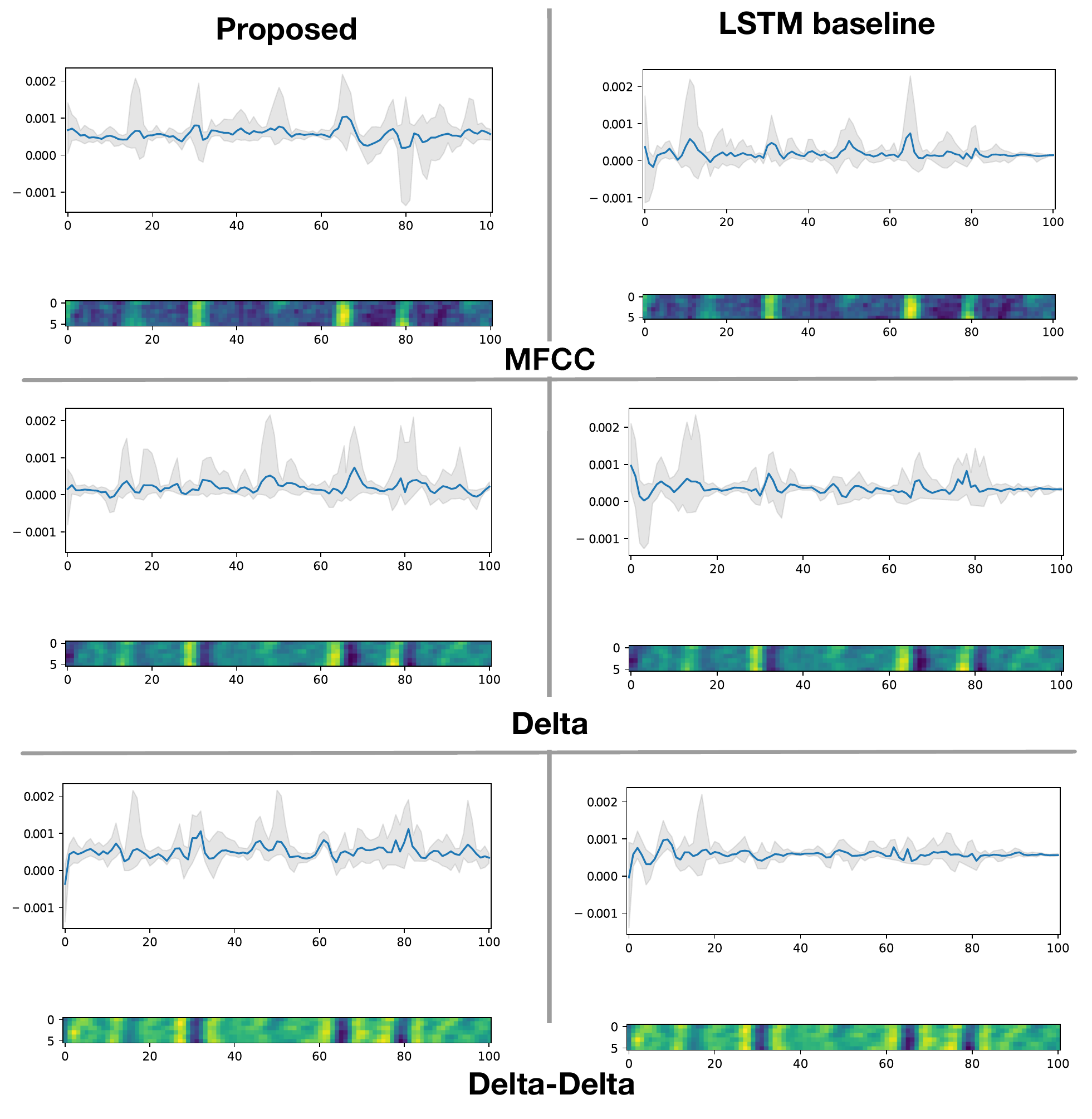}}
 \caption{SHAP \cite{lundberg2017unified, lundberg2018explainable} values denoting the change in the expected model predictions when the model is expected to generate a prediction of +1. In the rows we show MFCC, $\Delta$ and $\Delta^2$ features and in first and second columns we indicate proposed method and LSTM baselines. As MFCC, $\Delta$ and $\Delta^2$ features contain 6 coefficients, we visualise the mean value in blue and the distribution in grey. We also illustrate the heat map of the feature. We observe that the $\Delta$ and $\Delta^2$ features have impacted more towards generating accurate predictions from the proposed model compared to MFCC feature.}
\label{fig:shap_2}
\end{figure*}

Fig. \ref{fig:shap_1} illustrates the SHAP values for the proposed model together with the LSTM baseline when the model is expected to generate a prediction of -1, and Fig. \ref{fig:shap_2} shows when they are expected to generate a +1 prediction. As MFCC, $\Delta$ and $\Delta^2$  each contain 6 coefficients, we visualise the mean value in blue and the distribution in grey. We also provide the heat maps of each feature sample where yellow indicates high values for the feature and blue indicates low values. 

We observe that the $\Delta$ feature has more impact towards generating accurate predictions from the model compared to MFCC and $\Delta^2$ features, which seem to be effected by artefacts that are present in the recordings. Furthermore, after comparing the associated heat maps together with SHAP values we conclude that the predictions correspond to the heart sounds (denoted as yellow regions in the corresponding heat map) that are actually observed in the PCG. 

We further compare these SHAP values together with the SHAP values generated by the baseline LSTM system in the right column of Fig. \ref{fig:shap_1} and Fig. \ref{fig:shap_2}. It is clear that there are more fluctuations in the baseline LSTM model predictions and the model is unsure about the output that it produces. We observe much smoother predictions from the proposed model. 


\begin{figure*}[htbp]
\centering
 \subfloat[][Before training]{\includegraphics[width=.45\textwidth]{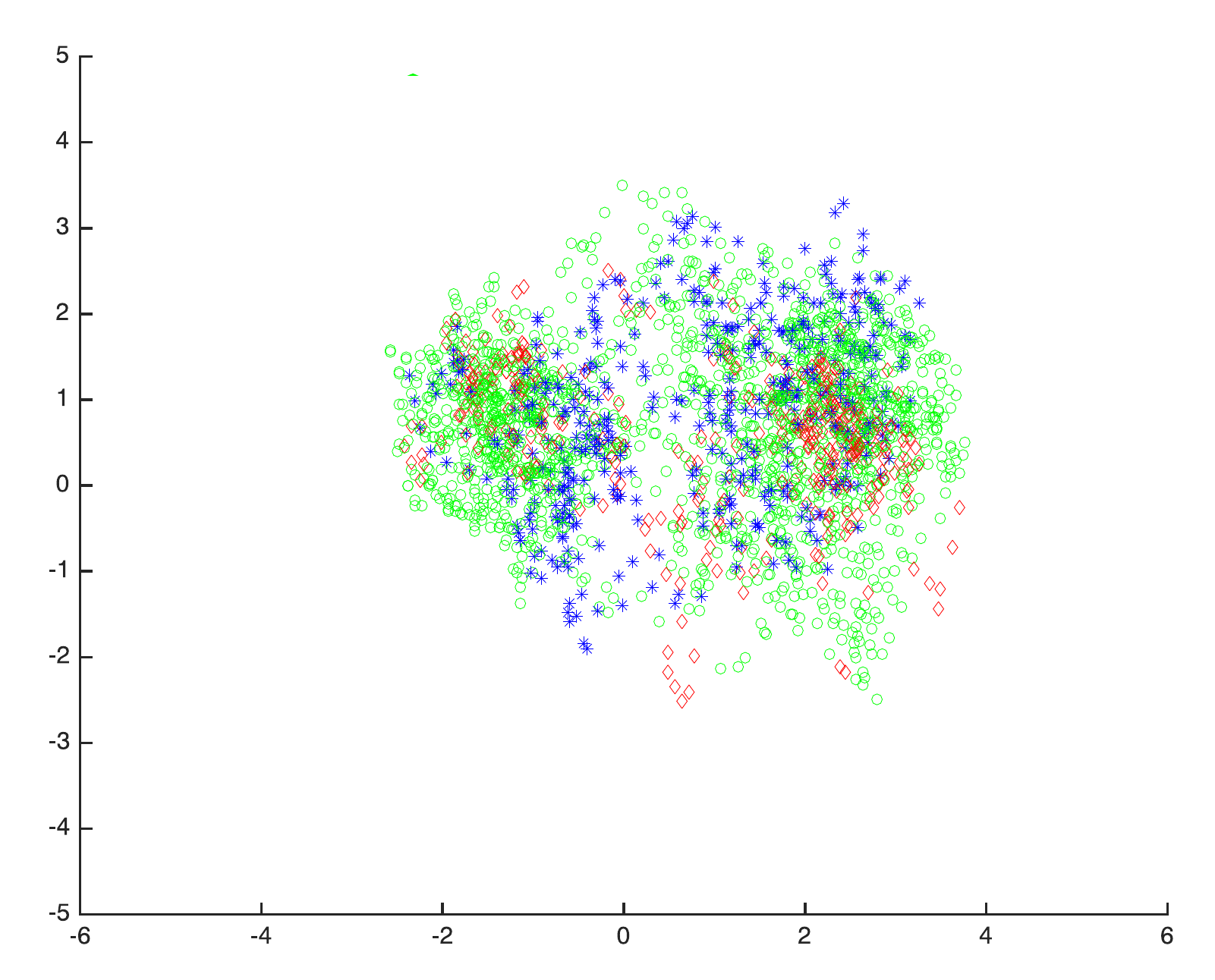}}
 \subfloat[][After training]{\includegraphics[width=.45\textwidth]{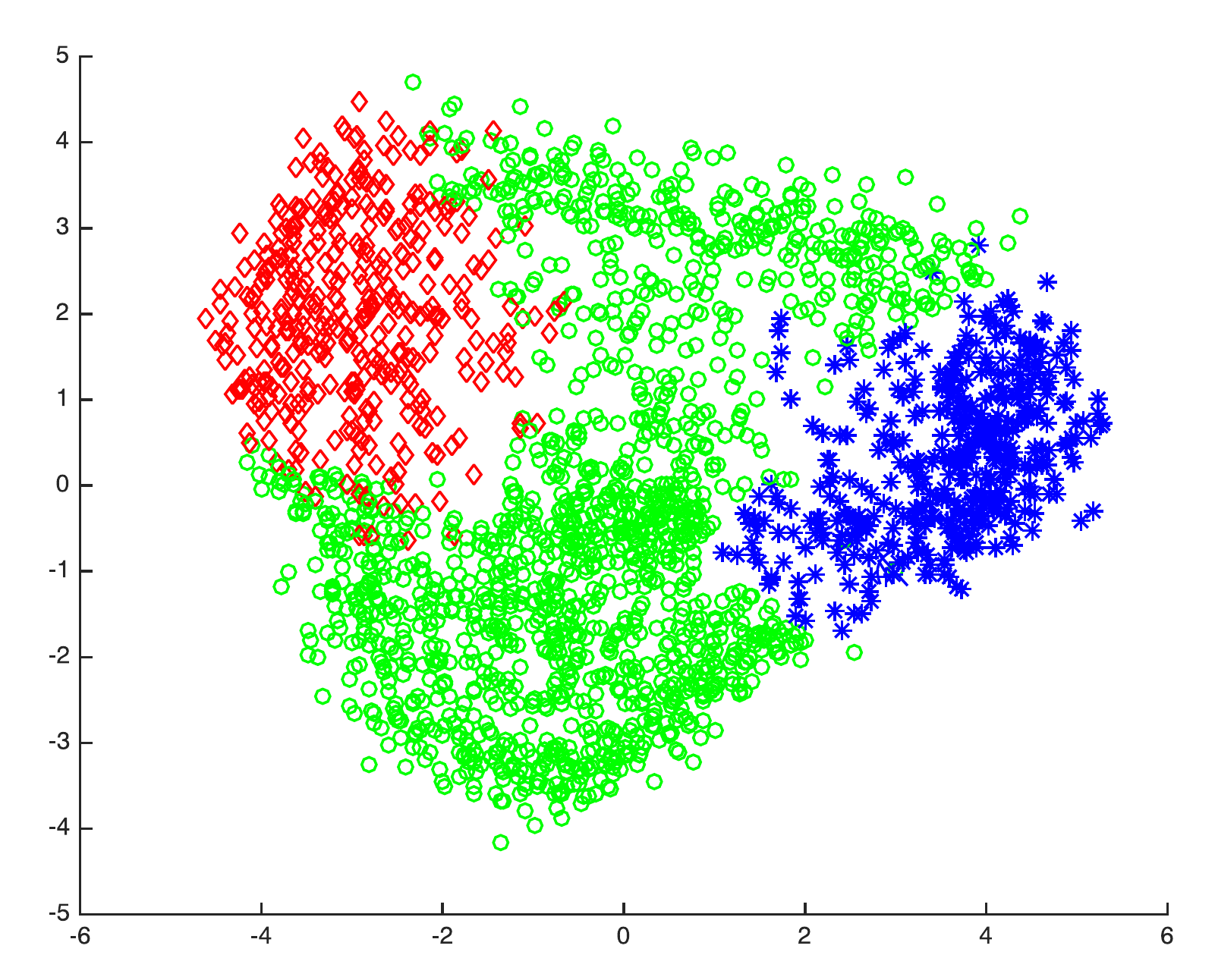}}
 \caption{ Projections of the output of Eq. \ref{eq:att}  before (a) and after (b) training. Red diamonds indicate ground truth heart state S1, a blue star indicates ground truth heart state S2 and a green circle indicates the None heart state.}
\label{fig:embedding_shift}
\end{figure*}

To further demonstrate the proposed segmentation model we analyse the discriminative nature of the learned embeddings from the proposed method. We selected 2500 examples from the M3-Hu dataset and visualise the output of Eq. \ref{eq:att} before training (Fig. \ref{fig:embedding_shift} (a)) and after training (Fig. \ref{fig:embedding_shift} (b)). Similar to \cite{fernando2018gd,aubakirova2016interpreting} we applied PCA \cite{wold1987principal} to plot the model outputs in 2D. In both figures we have colour coded the S1, S2 and None heart states. Red diamonds indicate ground truth heart state S1, blue stars indicate ground truth heart state S2 and a green circle indicates the None heart state.

Considering the examples given, it is clear that the model learns a representation from the features which better segregates the heart states. 

\subsection{Qualitative Results}

\begin{figure}[htbp]
\centering
 \subfloat[][]{\includegraphics[width=.48\linewidth]{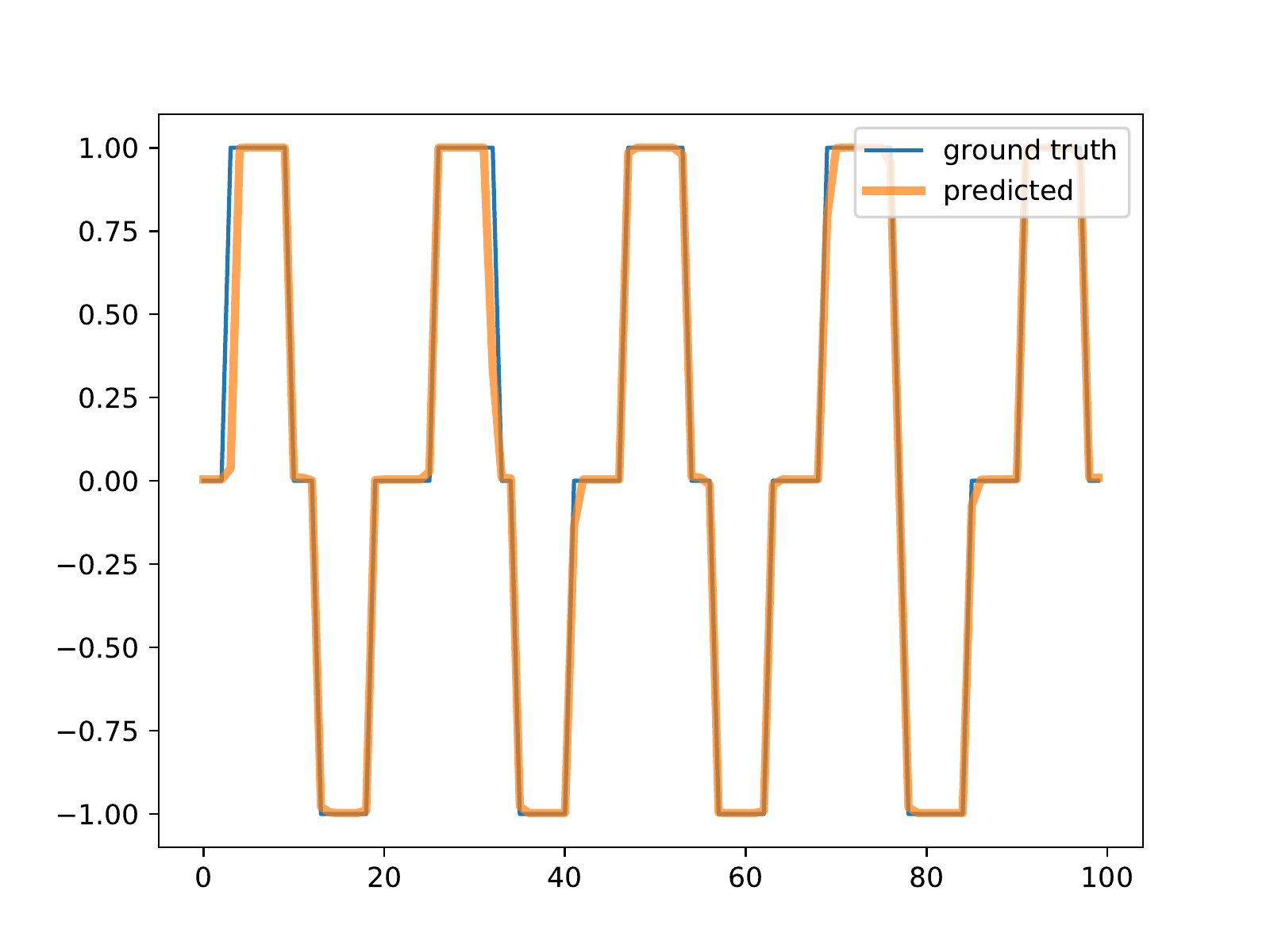}}
 \subfloat[][]{\includegraphics[width=.48\linewidth]{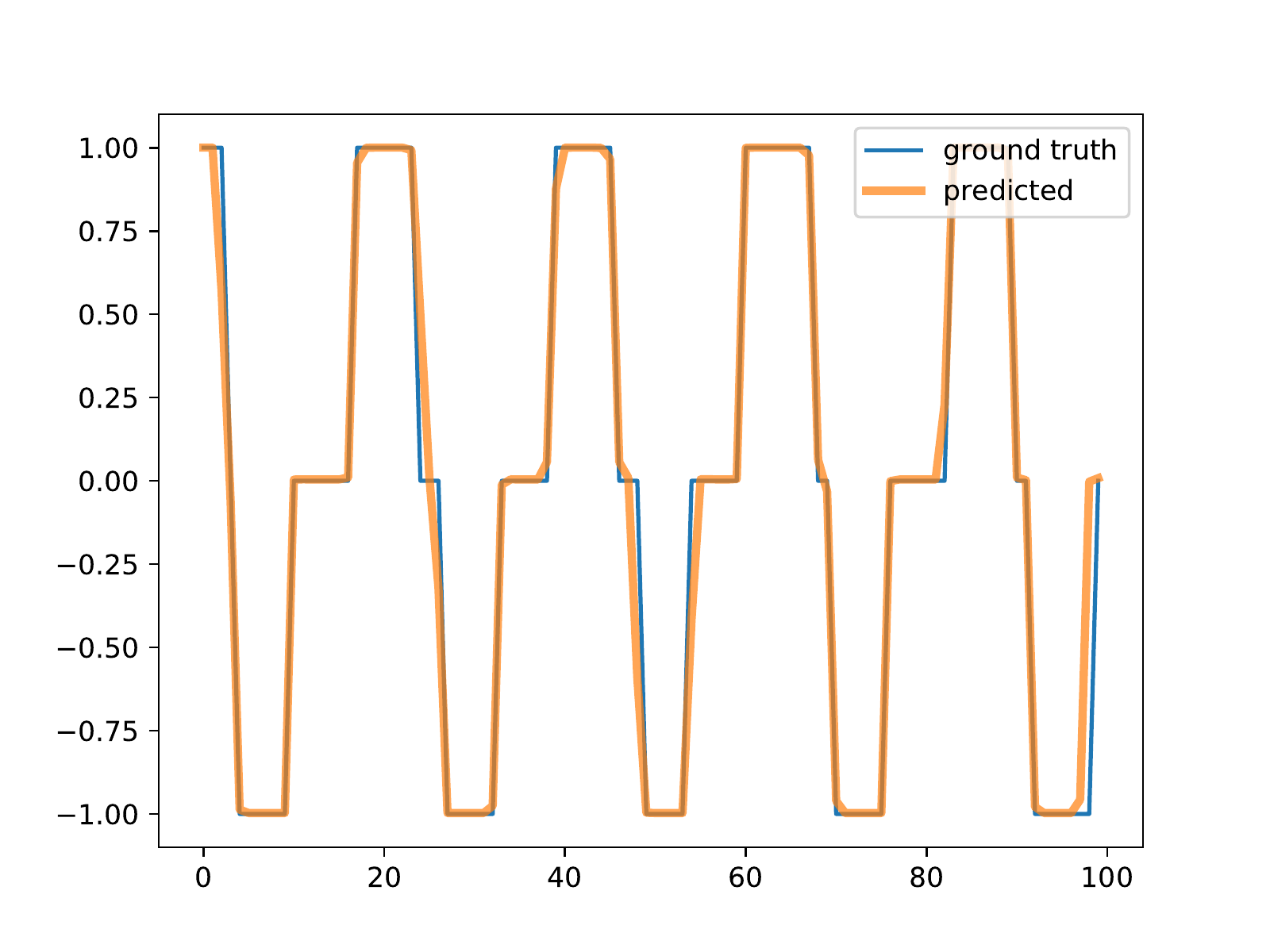}} \\
 \subfloat[][]{\includegraphics[width=.48\linewidth]{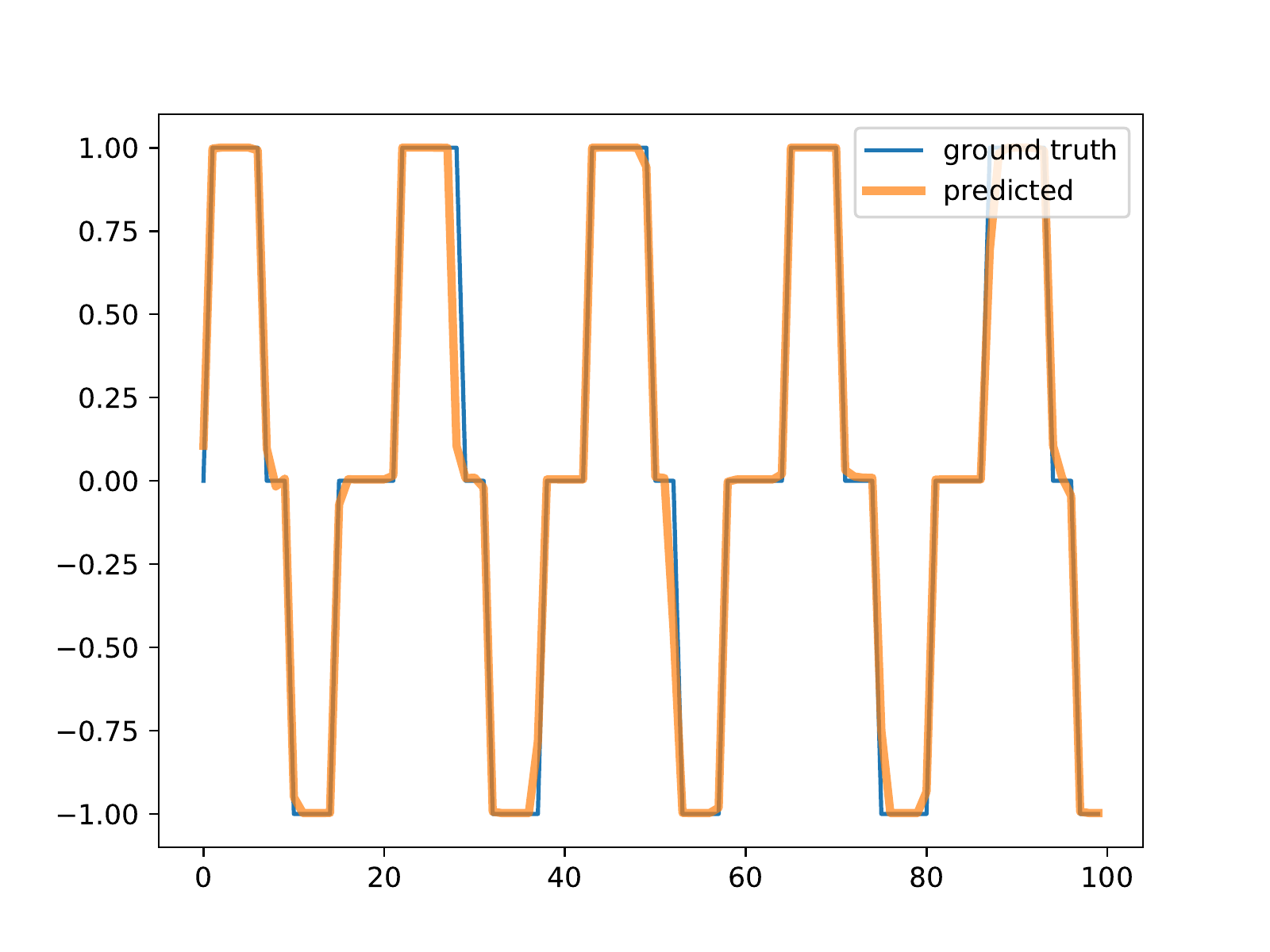}}
 \subfloat[][]{\includegraphics[width=.48\linewidth]{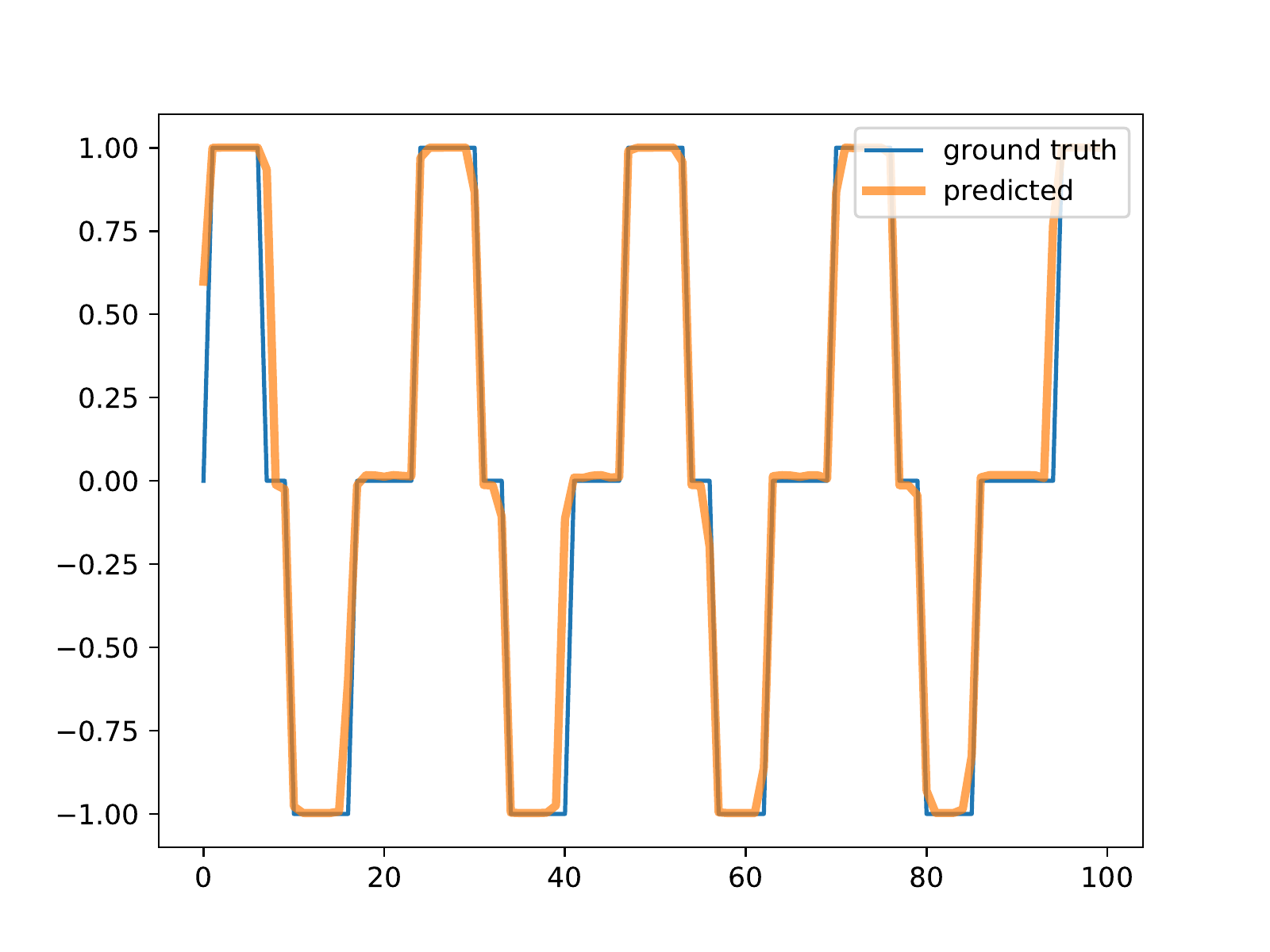}} \\
 \subfloat[][]{\includegraphics[width=.48\linewidth]{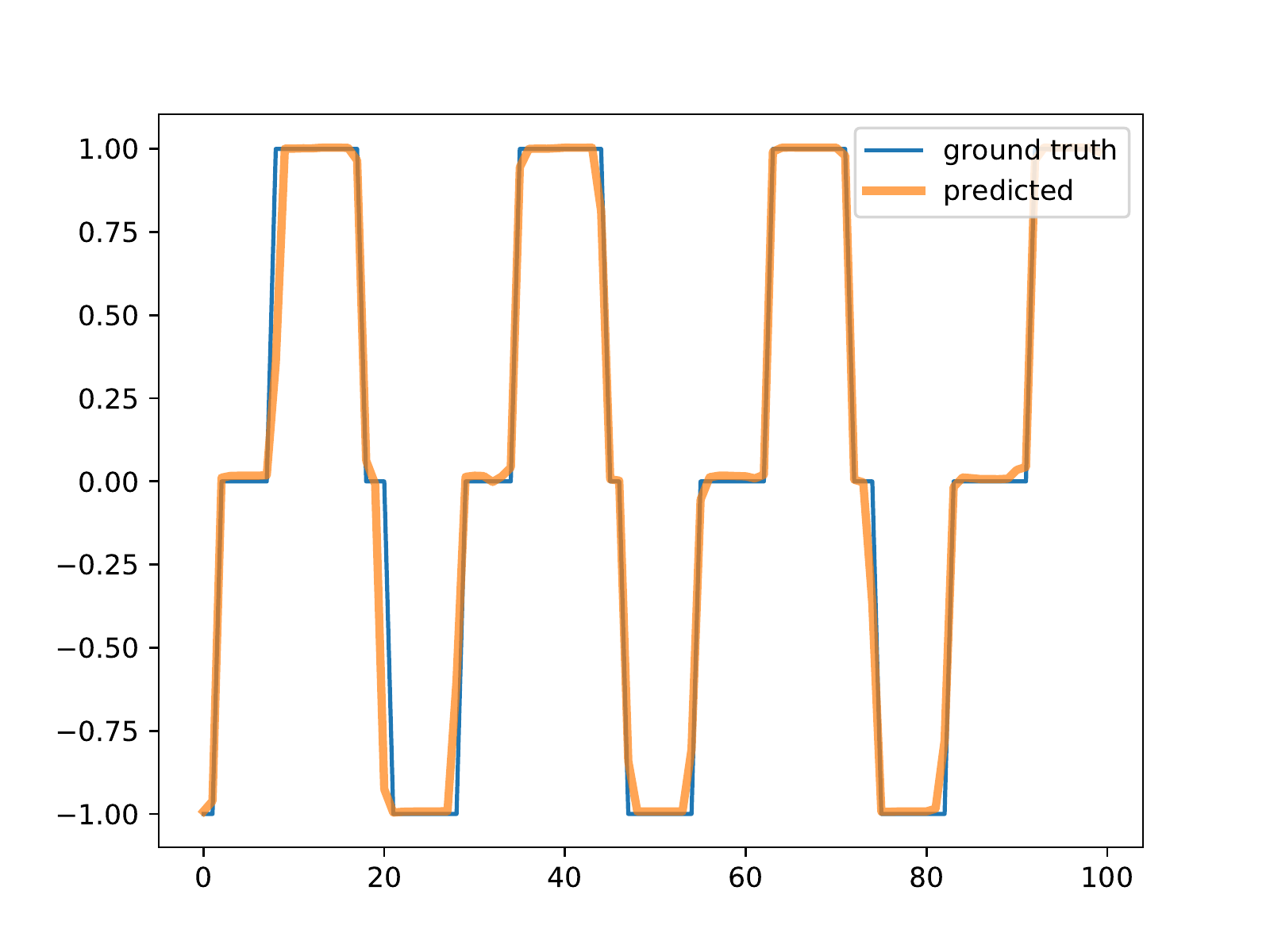}}
 \subfloat[][]{\includegraphics[width=.48\linewidth]{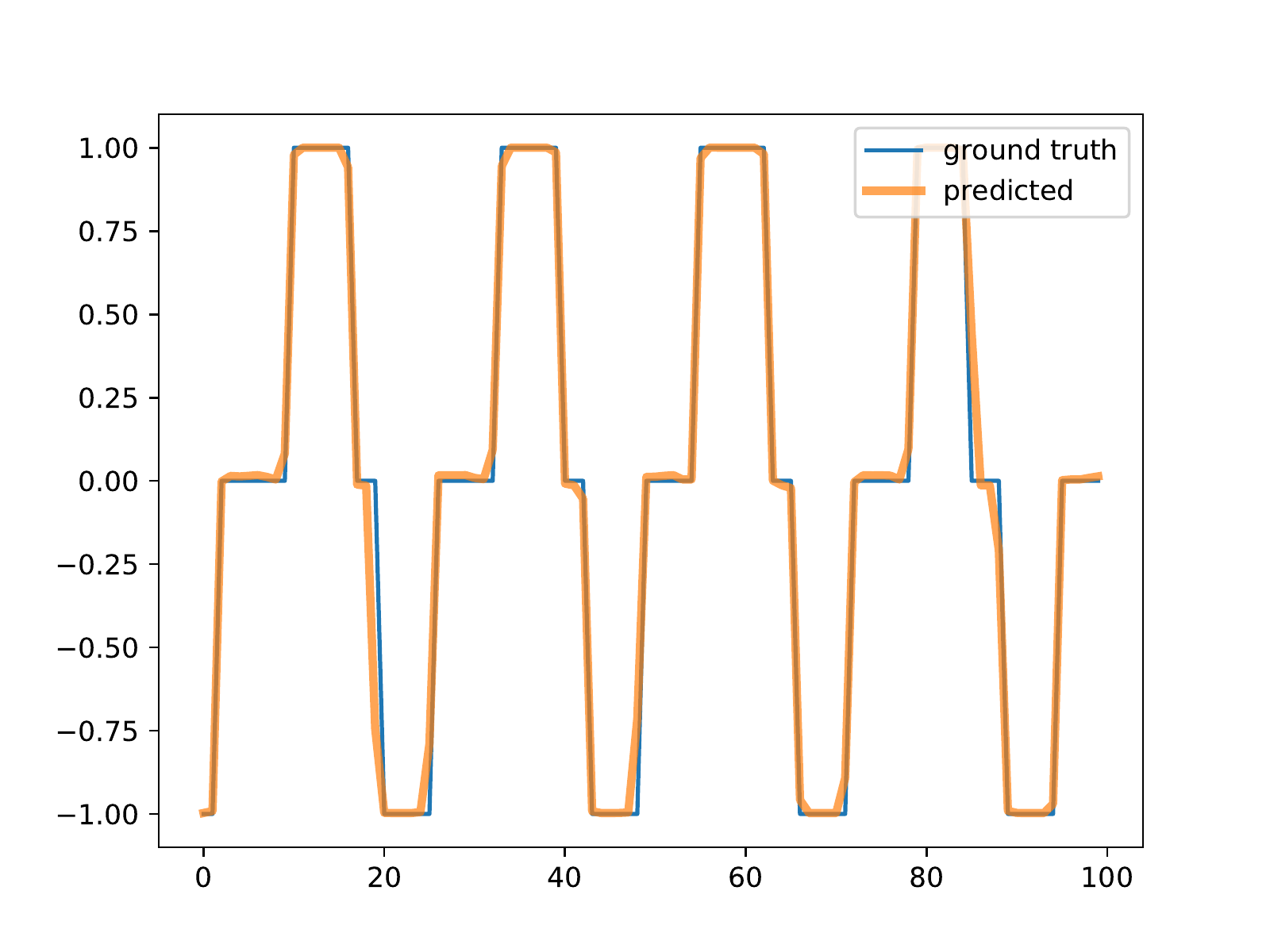}}
 \caption{Qualitative Results of automatically segmented heart sound recordings along with the ground truth annotations for M3-Hu dataset. We observe that predictions of the proposed method closely correspond to the ground truth.}
\label{fig:qualitative}
\end{figure}


In Fig. \ref{fig:qualitative} we visualise 6 examples of automatically segmented heart sound recordings along with the ground truth annotations for M3-Hu dataset. It is clear that the proposed method generates predictions that closely correspond to the ground truth. 

\subsection{Time Efficiency} 

The proposed model does not require any special hardware such as GPUs to run and has  17K trainable parameters, which is considerably less compared to the 52K parameters of the GRNN model in \cite{messner2018heart}. We ran the test set of M3-Hu on a single core of an Intel Xeon-2680 2.50 GHz CPU and the proposed model generates 1000 heart state classifications (S1, S2, None) in 56.88 seconds. In the same setting the method of \cite{messner2018heart} takes 124.45 seconds.

\section{Conclusion}
In this paper we introduce a novel deep learning framework for heart sound segmentation. We demonstrate how the recent successes in recurrent neural network based temporal modelling as well as attention based salient feature extraction techniques can be incorporated to mitigate the challenges posed by irregular and noisy PCG recordings. We conduct experiments with heart sound recordings from multiple benchmarks, including 2016 Physionet/CinC Challenge and privately collected data (M3dicine databases) that contains both animal and human heart sounds, where we outperform the current state-of-the-art methods. In addition, we quantitatively interpret the feature importance in the proposed architecture and further empirically analyse different feature combinations including envelope, wavelet and MFCC features. The evaluations demonstrate the utility of temporal modelling using recurrent neural networks and the effectiveness of the proposed attention mechanism in overcoming the challenges posed by noisy and irregular heart sound recordings. The proposed system not only achieves superior performance compared to state-of-the-art baseline methods, but is also significantly more efficient in terms of the number of trainable parameters and computational requirements during inference. The superior results that the proposed system attains in both the human and animal experimental settings affirms its utility in computer aided heart sound analysis and provides a platform for a variety of subsequent analysis and applications including detection of murmurs and ejection clicks. Furthermore, the proposed framework is not restricted for heart sound recordings and could be applied for analysing any one dimensional signal including lung sounds, electrocardiograms and electroencephalograms.


%

%


\section*{Acknowledgment}

The authors would like to thank Prof Anthony J. Sinskey from the Massachusetts Institute of Technology, Department of Biology for his assistance in some of the data collection and annotation outlined in this work. 
This research was supported by a Cooperative Research Centre Projects (CRC-P) grant. The authors also thank QUT High Performance Computing (HPC) for providing the computational resources for this research.

\ifCLASSOPTIONcaptionsoff
  \newpage
\fi



\bibliographystyle{IEEEtran}
\bibliography{egdb}

%

%






\end{document}